\documentclass[final,5p,times,twocolumn]{elsarticle}

\usepackage{amsmath}
\usepackage{amssymb}
\usepackage[normalem]{ulem}

\biboptions{sort&compress}

\usepackage[dvipsnames]{xcolor}
\usepackage{hyperref}
\hypersetup{
    colorlinks=true,
    linkcolor=Maroon,
    citecolor=MidnightBlue,
    urlcolor=red
}

\newcommand{\rg}{r_g}

\journal{Physics Letters A}

\begin{document}

\begin{frontmatter}

\title{Local Group Velocity Distribution inside Superradiant Condensates}

\author[a,b]{Yin-Da Guo}
\ead{yinda.guo@mail.sdu.edu.cn}

\author[a]{Kai-Dong Zhou}
\ead{kaidong.zhou@outlook.com}

\author[a]{Shou-Shan Bao}
\ead{ssbao@sdu.edu.cn}

\author[a]{Hong Zhang}
\ead{hong.zhang@sdu.edu.cn}

\affiliation[a]{organization={
                Key Laboratory of Particle Physics and Particle Irradiation (MOE), Institute of Frontier and Interdisciplinary Science},
            addressline={Shandong University},
            city={Qingdao},
            postcode={266237},
            state={Shandong},
            country={China}}
\affiliation[b]{organization={
                CENTRA, Departamento de Física, Instituto Superior Técnico - IST},
            addressline={Universidade de Lisboa - UL},
            city={Avenida Rovisco Pais 1},
            postcode={1049-001},
            state={Lisboa},
            country={Portugal}}

\begin{abstract}
Superradiance enables scalar fields to extract energy and angular momentum from a rotating black hole (BH), leading to the formation of a BH-condensate system. Previous studies mainly focus on the phase velocity, which propagates in the azimuthal direction. In this work, we show that the superradiant scalar condensate presents a nontrivial group velocity distribution. In the region sufficiently far from the BH, the condensate exhibits a radial velocity magnitude that approaches $ (\rg\mu/2) \sin (2\omega t-2 \varphi)$, while the polar and azimuthal velocity magnitudes asymptotically decline as $\propto 1/r$.
\end{abstract}


\end{frontmatter}




\section{Introduction}
\label{introduction}

\begin{figure}
	\centering 
	\includegraphics[width=0.4\textwidth]{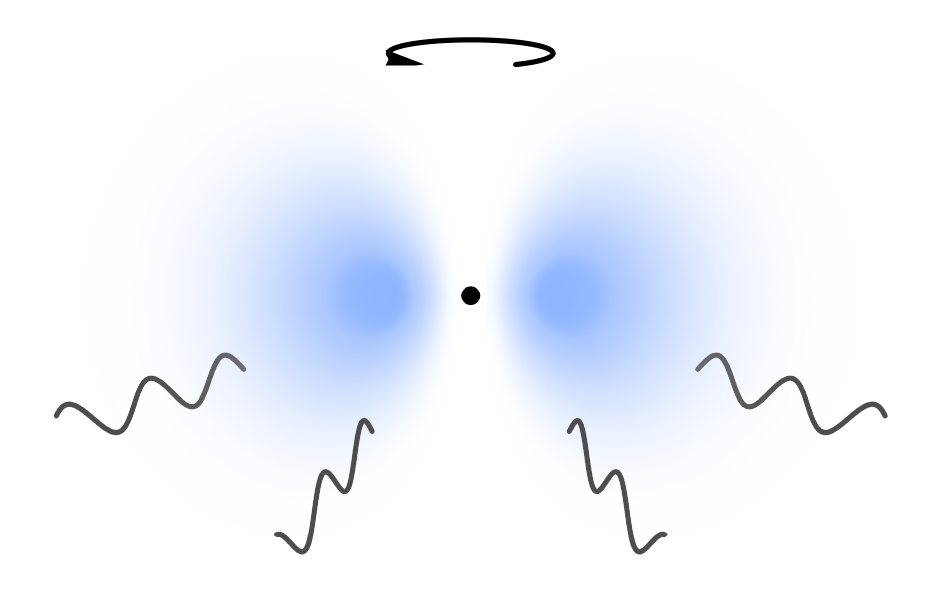}
    \caption{Schematic illustration of the BH-condensate system. The blue region represents the distribution of the real scalar field at a certain time. The black object at the center denotes the Kerr BH, which is enlarged by a factor of 10 for illustrative purposes. The system rotates and emits GWs due to the non-axisymmetry of the condensate. The GWs are depicted as gray wavy lines.}
	\label{fig:BH_condensate_schema}
\end{figure}

Ultra-light bosons can extract energy and angular momentum from a rotating black hole (BH) via the superradiance mechanism \cite{Press:1972zz,Misner:1972kx}. This mechanism occurs when the frequency $\omega$ of the bosons satisfies the superradiant condition, $\omega < m \Omega_\mathrm{H}$, where $\Omega_\mathrm{H}$ is the angular velocity of the event horizon and $m$ is the magnetic number of the boson. The transferred energy can reach up to approximately 10\% of the mass of the BH \cite{East:2017ovw,Herdeiro:2017phl,Guo:2025dkx}. Unlike fermions, these bosons can condense in large numbers without being subject to the Pauli exclusion principle, thereby forming BH-condensate systems, schematically illustrated in Fig.~\ref{fig:BH_condensate_schema}. A detailed account of the BH superradiance is provided in Ref.~\cite{Brito:2015oca}.

Superradiance has been studied in detail in the literature for scalar \cite{Arvanitaki:2010sy,Yoshino:2013ofa,Yoshino:2014wwa,Arvanitaki:2014wva,Arvanitaki:2016qwi,Brito:2017wnc,Brito:2017zvb,Herdeiro:2017phl,LIGOScientific:2021rnv,Sun:2019mqb,Guo:2022mpr,Guo:2025dkx,Fernandez:2019qbj,Ng:2019jsx,Ng:2020ruv,Cheng:2022jsw,Brito:2014wla,Fukuda:2019ewf,Roy:2021uye,Hui:2022sri,Sarmah:2024nst,Unal:2023yxt,Yoshino:2012kn,Dolan:2012yt,Baryakhtar:2020gao,Omiya:2022mwv,Omiya:2022gwu,Omiya:2024xlz,Xie:2025npy,Baumann:2018vus,Baumann:2021fkf,Baumann:2022pkl,Tong:2022bbl,Takahashi:2023flk,Takahashi:2024fyq,Brito:2023pyl,Zouros:1979iw,Blas:2020nbs,Chen:2019fsq,Chen:2021lvo,Chen:2022oad,Detweiler:1980uk,Cardoso:2005vk,Konoplya:2006br,Dolan:2007mj,Arvanitaki:2009fg,Konoplya:2011qq,Yoshino:2015nsa,Ficarra:2018rfu,Tong:2021whq,Bao:2022hew,Chen:2022kzv,Yuan:2022nmu,Bao:2023xna,Yang:2023vwm,Dai:2023ewf,Dolan:2024qqr,Chu:2024iie,Zhu:2024bqs,DellaMonica:2025zby,Jia:2025vqn,Witek:2012tr,Endlich:2016jgc,Baumann:2019eav,Cardoso:2018tly,Isi:2018pzk},
vector \cite{Cardoso:2018tly,Herdeiro:2017phl,Baryakhtar:2017ngi,East:2018glu,Siemonsen:2019ebd,Jones:2024fpg,Mirasola:2025car,Guo:2024dqd,Rosa:2011my,Witek:2012tr,Pani:2012vp,Pani:2012bp,Endlich:2016jgc,Dolan:2018dqv,East:2017mrj,East:2017ovw,Frolov:2018ezx,Isi:2018pzk,Baumann:2019eav,Percival:2020skc,Caputo:2021efm,East:2022ppo,Jia:2023see,May:2024npn,Jia:2025vqn,Chen:2022nbb}
and tensor \cite{Chen:2022nbb,Brito:2013wya,Brito:2020lup,Dias:2023ynv,East:2023nsk} fields. 
The simplest case involves studying a free bosonic field in a Kerr spacetime. The resulting BH-condensate system can emit potentially observable gravitational waves (GWs), including 
quasi-monochromatic continuous GWs \cite{Arvanitaki:2010sy,Yoshino:2013ofa,Yoshino:2014wwa,Arvanitaki:2014wva,Arvanitaki:2016qwi,Brito:2017wnc,Brito:2017zvb,Baryakhtar:2017ngi,East:2018glu,Isi:2018pzk,Siemonsen:2019ebd,Sun:2019mqb,Brito:2020lup,LIGOScientific:2021rnv,Guo:2022mpr,Jones:2024fpg,Guo:2024dqd,Guo:2025dkx,Mirasola:2025car}, 
stochastic GWs \cite{Brito:2017wnc,Brito:2017zvb} and GW beats \cite{Siemonsen:2019ebd,Guo:2022mpr,Guo:2024dqd,Guo:2025dkx}. The formation of these systems also leads to a distinctive gap in the Regge plane—a plot that illustrates the relationship between BH spin and mass. Consequently, by measuring the spins and masses of a large number of BHs, the presence of BH-condensate systems can be indirectly determined or constrained \cite{Arvanitaki:2010sy,Cardoso:2018tly,Fernandez:2019qbj,Ng:2019jsx,Ng:2020ruv,Cheng:2022jsw,Guo:2024dqd}. 

Beyond the simplest scenario, more complex situations have been investigated. Taking accretion into account, the condensate mass can grow to as much as 20\% of the BH mass, thereby significantly enhancing the GW signals while reducing their duration \cite{Brito:2014wla,Fukuda:2019ewf,Roy:2021uye,Hui:2022sri,Unal:2023yxt,Sarmah:2024nst,Guo:2025dkx}. Bosons may interact with themselves or with the Standard Model particles. For instance, the scalar-photon interaction could modify the cosmic microwave background \cite{Blas:2020nbs} and lead to a birefringent effect \cite{Chen:2019fsq,Chen:2021lvo,Chen:2022oad}. In addition, the scalar self-interaction could result in a bosenova collapse or suppress the growth of condensate \cite{Yoshino:2012kn, Fukuda:2019ewf,Baryakhtar:2020gao,Omiya:2022mwv,Omiya:2022gwu,Unal:2023yxt,Omiya:2024xlz,Xie:2025npy}, thereby altering the expected GW signal. Moreover, when the BH-condensate system resides in a binary, energy level transitions or ionization may occur, which in turn modifies the GW signals of the binary \cite{Baumann:2018vus,Baumann:2021fkf,Baumann:2022pkl,Tong:2022bbl,Takahashi:2023flk,Brito:2023pyl,Takahashi:2024fyq}.

In this work, we study the simplest scalar scenario, focusing on the local group velocity distribution of the scalar field.
This distribution must be very different from that of the phase velocity. Otherwise the region of the condensate far from the BH would be superluminal, which is clearly unphysical.
Apart from the theoretical interest to find out what is the local motion of the scalar field leading to the apparent rotation of the condensate, the group velocity distribution is also phenomenologically relevant.
Taking the photon traversing the condensate as an example, 
it is the relative motion of the photon and the local scalar field which determines the photon path and the accumulated phase shift.
This effect could be important, especially when the group velocity is close to the speed of light.
The purpose of this paper is to calculate the precise group velocity distribution of a superradiant scalar condensate. We leave the study of its phenomenological consequence in a future work.

This paper is organized as follows. In Sec.~\ref{sec:BHCondensate}, we briefly review scalar superradiance and the BH-condensate system. Then the normalization of the scalar field is discussed in Sec.~\ref{sec:normalization}. In Sec.~\ref{sec:Rotation}, we examine the local group velocity distribution in the condensate. Finally, we present our summary and discussion in Sec.~\ref{sec:Summary}.

Throughout the paper, we adopt the natural units $\hbar=c=1$.

\section{Black hole-condensate system}
\label{sec:BHCondensate}

The Kerr metric, describing a rotating BH with mass $M$ and angular momentum $J$, is given in Boyer-Lindquist coordinates as follows \cite{Boyer:1966qh}:
\begin{align}
    \begin{split}
        ds^2=&\left( 1-\frac{2 \rg r}{\Sigma} \right) dt^2+\frac{4a \rg r}{\Sigma}\sin ^2\theta dtd\varphi -\frac{\Sigma}{\Delta}dr^2
        \\
        &-\Sigma d\theta ^2-\left[ \left( r^2+a^2 \right) \sin ^2\theta +2\frac{\rg r}{\Sigma}a^2\sin ^4\theta \right] d\varphi ^2,
    \label{eq:KerrMetric}
    \end{split}
\end{align}
where
\begin{subequations}
    \begin{align}
        a & \equiv J/M,\\
        \rg & \equiv GM, \\
        \Delta & \equiv r^2-2 \rg r+a^2,\\
        \Sigma & \equiv r^2+a^2 \cos ^2 \theta.
    \end{align}
\end{subequations}
Here, $G$ denotes the gravitational constant and the parameter $a$ corresponds to the angular momentum per unit BH mass. It is conventional to introduce a dimensionless parameter $a_* \equiv a/\rg$ and refer to it as the BH {\it spin}. The BH has an inner horizon $r_-$ and an outer horizon $r_+$, which are defined as
\begin{align}
    r_{ \pm}=\rg \pm \sqrt{\rg^2-a^2}.
\end{align}
Choosing a locally non-rotating frame, the orthonormal basis vectors have the following form \cite{Bardeen:1972fi}:
\begin{subequations}
    \begin{align}
        \left(e_{t}\right)^{\nu} & =\left(\frac{\Sigma\Delta}{A}\right)^{-1/2}\left[\left(\frac{\partial}{\partial t}\right)^{\nu}+\Omega\left(\frac{\partial}{\partial\varphi}\right)^{\nu}\right],
        \\
        \left(e_{r}\right)^{\nu} & =\left(\frac{\Sigma}{\Delta}\right)^{-1/2}\left(\frac{\partial}{\partial r}\right)^{\nu},
        \\
        \left(e_{\theta}\right)^{\nu} & =\Sigma^{-1/2}\left(\frac{\partial}{\partial\theta}\right)^{\nu},
        \\
        \left(e_{\varphi}\right)^{\nu} & =\left(\frac{A\sin^{2}\theta}{\Sigma}\right)^{-1/2}\left(\frac{\partial}{\partial\varphi}\right)^{\nu}.
    \end{align}
\end{subequations}
where $\Omega\equiv 2ar_{g}r/A$ and $A \equiv \left(r^{2}+a^{2}\right)^{2}-\Delta a^{2}\sin^{2}\theta$. Their dual vectors are
\begin{align}
    \left(e^{t}\right)_{\nu} & =\left(\frac{\Sigma\Delta}{A}\right)^{1/2}\left(dt\right)_{\nu},\\
    \left(e^{r}\right)_{\nu} & =\left(\frac{\Sigma}{\Delta}\right)^{1/2}\left(dr\right)_{\nu},\\
    \left(e^{\theta}\right)_{\nu} & =\Sigma^{1/2}\left(d\theta\right)_{\nu},\\
    \left(e^{\varphi}\right)_{\nu} & =\left(\frac{A\sin^{2}\theta}{\Sigma}\right)^{1/2}\left[-\Omega\left(dt\right)_{\nu}+\left(d\varphi\right)_{\nu}\right].
\end{align}
In this basis, the metric tensor reduces to the diagonal form $\{1, -1, -1, -1\}$.

In this work, we investigate a real scalar field bound in the background of a Kerr BH. Given that the energy density of the condensate is much lower than that of the BH, its backreaction on the spacetime geometry can be safely neglected \cite{Brito:2014wla}. We also disregard the self-interaction and interactions with other fields. Under these assumptions, the scalar field then obeys the Klein-Gordon equation
\begin{align}
	\left(\nabla^{\nu}\nabla_{\nu}+\mu^{2}\right)\Phi=0,
	\label{eq:KG_equation}
\end{align}
where $\mu$ is the scalar mass. The corresponding solution can be expressed as 
\begin{align}\label{eq:real_scalar}
	\Phi(t,r,\theta,\varphi) = \sum_{l,m} \int d\omega \frac{f_{lm}(\omega)}{\sqrt{2\omega}}\left[\phi_{lm}(t,r,\theta,\varphi)+\phi^*_{lm}(t,r,\theta,\varphi)\right],
\end{align}
where $\omega$, $l$ and $m$ denote the eigenfrequency, azimuthal number and magnetic number, respectively. In addition, $f_{lm}$ is the distribution function, and ``*'' denotes complex conjugation. The $\phi_{lm}$ also satisfies Klein-Gordon equation~\eqref{eq:KG_equation}. By adopting the following ansatz
\begin{align}
	\phi_{lm}(t,r,\theta,\varphi) = e^{-i \omega t}e^{i m \varphi} R_{lm}(r)S_{lm}(\theta),
\end{align}
the Klein-Gordon equation~\eqref{eq:KG_equation} can be separated into radial and angular parts. The resulting equations take the form
\begin{subequations}
    \begin{align}
        \label{eq:radial_equation}
	    \begin{split}
            \Delta \frac{d}{d r}\left(\Delta \frac{d R_{l m}(r)}{d r}\right)+\Bigg[\omega^{2}\left(r^{2}+a^{2}\right)^{2}- 4 m a \rg \omega r  +a^{2} m^{2} &
            \\
            -\Delta\left(\mu^{2} r^{2}+a^{2} \omega^{2}+\Lambda_{l m}\right) \Bigg] R_{l m}(r) &=0,
        \end{split}
        \\
        \label{eq:angular_equation}
        \begin{split}
            \frac{1}{\sin \theta} \frac{d}{d \theta}\left(\sin \theta \frac{d S_{l m}(\theta)}{d \theta}\right) +\Bigg[-a^{2}\left(\mu^{2}-\omega^{2}\right) \cos ^{2} \theta &
            \\
            -\frac{m^{2}}{\sin ^{2} \theta}+\Lambda_{l m}\Bigg] S_{l m}(\theta) &=0.
        \end{split}
    \end{align}
\end{subequations}
The angular equation \eqref{eq:angular_equation} is the eigenequation for the spheroidal harmonics $S_{l m}(\theta)$, with $\Lambda_{l m}$ denoting the corresponding eigenvalue. 

The radial function and the eigenfrequency can be obtained by solving the radial equation~\eqref{eq:radial_equation}, either perturbatively in the $\rg \mu \ll 1$ limit following the method invented by Detweiler~\cite{Detweiler:1980uk} and later developed in Refs.~\cite{Pani:2012bp,Bao:2022hew,Bao:2023xna}, or numerically using the continued fraction method \cite{Leaver:1985ax,Cardoso:2005vk,Dolan:2007mj}. We impose the boundary conditions for quasi-bound states, requiring the radial function to vanish at spatial infinity and to be purely ingoing near the horizon. 

Generally, the eigenfrequencies are complex numbers and are characterized by three indices $\{n,l,m\}$, with $n$ denoting the overtone number. Their possible values are $n=0,1,2,\ldots$, $l=1,2,3,\ldots$, and $m\in[-l,l]$. The principal number $\bar{n}=n+l+1$ is also commonly used in place of $n$. We denote the real and imaginary parts of the eigenfrequency as $\omega_{nlm}$ and $\Gamma_{nlm}$, respectively. In the non-relativistic limit, the solutions of Eq.~\eqref{eq:KG_equation} take a hydrogen-like form, and the real part of the eigenfrequency is given by
\begin{align}\label{eq:NR-omega}
	\omega_{nlm} = \mu \left[1-\frac{\left(\rg\mu\right)^2}{2(n+l+1)^2} + \mathcal{O}\left(\rg^4\mu^4\right)\right].
\end{align}
Higher-order corrections to the real part can be found in Ref.~\cite{Baumann:2019eav}. The imaginary part is also referred to as the superradiant rate. A positive value of $\Gamma_{nlm}$ indicates energy transfer from the BH to the scalar field, consistent with the superradiant condition $\omega<m\Omega_\mathrm{H}$. Here, $\Omega_\mathrm{H} \equiv a/(2r_gr_+)$ is the angular velocity of the event horizon. In this case, the scalar field grows exponentially, leading to the formation of a BH-condensate system, as illustrated in Fig.~\ref{fig:BH_condensate_schema}. If $\Gamma_{nlm}<0$, energy flows from the scalar field to the BH.


After obtaining the eigenfrequency, the radial function $R(r)$ can be determined by \cite{Dolan:2007mj}
\begin{align}
	R_{nlm}(r)=(r-r_+)^{-i\sigma}(r-r_-)^{i\sigma+\chi-1}e^{qr}\sum_{k=0}^\infty a_k{\left(\frac{r-r_+}{r-r_-}\right)^k},
\end{align}
with
\begin{align}
    \sigma &= \frac{2r_+(\omega_{nlm}-m\Omega_\mathrm{H})}{r_+-r_-}, 
    \\
    q &= -\sqrt{\mu^2-\omega^2_{nlm}},
    \\
    \chi &= \frac{\mu^2-2\omega^2_{nlm}}{q}.
\end{align}
Here, the subscript $n$ is added to $R$ to show its dependence on $\omega_{nlm}$. The functions $S$ and $\phi$ will be similarly relabeled below. In addition, the coefficients $a_k$ can be obtained from the recurrence relation provided in Ref.~\cite{Dolan:2007mj}. The peak position of the radial function can be denoted by $r_\mathrm{p}$ and the length scale can be characterized by the Bohr radius $r_\mathrm{b} = 1/(r_g \mu^2)$.

The modes in the condensate with different values of $l = m$ result in distinct evolution stages \cite{Ficarra:2018rfu,Guo:2022mpr}. Modes with smaller $l = m$ have larger superradiant rates and GW emission fluxes. Thus, they experience earlier and shorter evolution stages. The modes with $n = 0$ are typically the dominant modes in each stage. These $n = 0$ modes initially grow exponentially until the BH spin drops to the values when the superradiance threshold $\omega = m \Omega_H$ is reached. Their masses gradually decrease due to GW emission thereafter, with a large fraction dissipated at the GW emission timescale.
In this work, we focus on the most unstable mode $\{0,1,1\}$ after it reaches the superradiance threshold, and examine its local group velocity distribution. 

\section{Normalization}
\label{sec:normalization}

In this section, we discuss the normalization of $\phi_{nlm}$, which is crucial for establishing the relationship between the coefficient $f_{lm}$ in Eq.~\eqref{eq:real_scalar} and the total energy of the condensate. We first analyze the case in flat spacetime and then extend the discussion to the general cases.

We begin with the stress-energy tensor for a real scalar field, which reads
\begin{align}\label{eq:stress-energy_tensor}
    T_{\nu\sigma} = \partial_{\nu}\Phi \partial_{\sigma}\Phi -g_{\nu\sigma}\left(\frac{1}{2}\partial_{\delta}\Phi \partial^{\delta}\Phi -\frac{1}{2}\mu^{2}\Phi^2\right).
\end{align}
Thus, the energy density $\rho$ of the field can be defined by
\begin{align}\label{eq:energy_density}
    \rho = T^{\nu}_{\ \sigma} \left(e_{t}\right)^{\sigma} \left(e^{t}\right)_{\nu}.
\end{align}
The corresponding total energy $M_\mathrm{s}$ of the scalar field is then given by 
\footnote{When self-gravity is taken into account, Eq.~\eqref{eq:Ms} does not represent the total energy, and the Arnowitt-Deser-Misner mass should instead be used.}
\begin{align}\label{eq:Ms}
    M_\mathrm{s} \equiv \int \rho \sqrt{|g|} dr d\theta d\varphi,
\end{align}
where $g$ is the determinant of the metric, and for Kerr spacetime, $\sqrt{|g|}=(r^2+a^2\cos^2\theta)\sin\theta$.

In flat spacetime, the total energy reduces to
\begin{align}
    M_\mathrm{s}^\mathrm{(flat)} = \int \frac{1}{2}
    \left[\partial_{t}\Phi\partial^{t}\Phi
    -\Phi\partial_{t}\partial^{t}\Phi
    -\nabla_{i}\left(\Phi\partial^{i}\Phi\right)\right]
    \sqrt{|g|} dr d\theta d\varphi,
\end{align}
where $i$ denotes spatial components and $\nabla$ represents the covariant derivative. If the scalar field is in a bound state, the third term of the integral vanishes and the corresponding eigenfrequency becomes discrete. Thus, $f_{lm}(\omega)$ can be rewritten by $\bar{f}_{nlm} \delta(\omega-\omega_{nlm})$ in Eq.~\eqref{eq:real_scalar}. Substituting Eq.~\eqref{eq:real_scalar} into the integral and considering only a single mode, the total energy becomes 
\begin{align}\label{eq:Ms_flat}
    M_\mathrm{s}^\mathrm{(flat)} = \bar{f}_{nlm}^2 \omega_{nlm} \int \left|\phi_{nlm}\right|^2 \sqrt{|g|} dr d\theta d\varphi. 
\end{align}
This indicates $M_\mathrm{s} = \bar{f}_{nlm}^2 \omega_{nlm}$ in flat spacetime when $\phi_{nlm}$ satisfies the normalization condition
\begin{align}\label{eq:normalization}
    \int \left|\phi_{nlm}\right|^2 \sqrt{|g|} dr d\theta d\varphi = 1.
\end{align}

\begin{figure}[t]
	\centering 
	\includegraphics[width=0.4\textwidth]{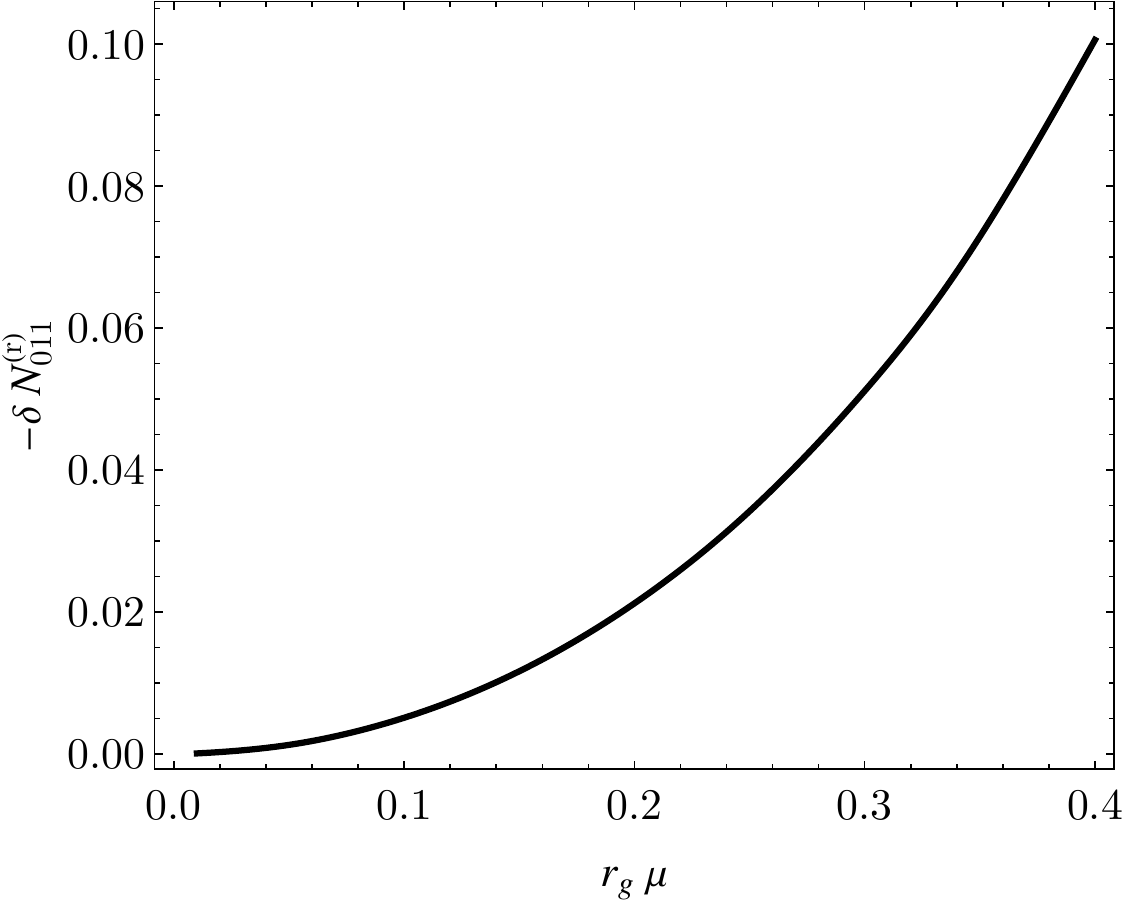}	
	\caption{The relative difference $\delta N^{(\mathrm{r})}_{011}$ defined in Eq.~\eqref{eq:delta_N} as a function of the mass coupling $\rg\mu$. Only the $\{0,1,1\}$ mode is considered. $\phi_{011}$ is numerically calculated as described in Sec.~\ref{sec:BHCondensate} and satisfies the normalization condition~\eqref{eq:normalization}. The BH spin $a_*$ is set to the value that satisfies the superradiant threshold $\omega_{011} = \Omega_\mathrm{H}$ for each value of $\rg \mu$.} 
	\label{fig:particle_number_vs_Mmu}
\end{figure}

However, $M_\mathrm{s}$ is not equal to $\bar{f}_{nlm}^2 \omega_{nlm}$ in curved spacetime even if $\phi_{nlm}$ satisfies the normalization condition~\eqref{eq:normalization}. For example, in the Schwarzschild metric, Eq.~\eqref{eq:Ms} reduces to
\begin{align}
    \begin{split}
        M_\mathrm{s}^\mathrm{(Schw.)} = \int 
        \Bigg[ & \bar{f}_{nlm}^2 \omega_{nlm} \frac{r}{r-2\rg}\left|R_{nlm}(r)S_{nlm}(\theta)\right|^2 e^{2\Gamma_{nlm}t} 
        \\
        & -\frac{1}{2}\nabla_{i}\left(\Phi\partial^{i}\Phi\right)\Bigg]
        \sqrt{|g|} dr d\theta d\varphi.  
    \end{split}
\end{align}
The total derivative term is non-zero since the event horizon is a dissipative surface, but it can still be neglected since the condensate is concentrated around $r \sim r_\mathrm{b} \gg \rg$. The additional factor $r/(r - 2\rg)$ in the first term leads to the deviation of $\bar{f}_{nlm}^2$ from $M_\mathrm{s}/\omega_{nlm}$.

Figure~\ref{fig:particle_number_vs_Mmu} shows the relative difference $\delta N^{(\mathrm{r})}_{nlm}$ between $\bar{f}_{nlm}^2$ and $M_\mathrm{s}/\omega_{nlm}$ for the $\{0,1,1\}$ mode in Kerr spacetime, defined as
\begin{align}\label{eq:delta_N}
    \delta N^{(\mathrm{r})}_{nlm} \equiv  \frac{\bar{f}_{nlm}^2}{M_\mathrm{s}/\omega_{nlm}} - 1,
\end{align}
where $\phi_{nlm}$ is numerically calculated as described in Sec.~\ref{sec:BHCondensate} with normalization~\eqref{eq:normalization}. It is apparent that the relation $M_\mathrm{s} = \bar{f}_{nlm}^2 \omega_{nlm}$ holds approximately only in the non-relativistic limit, $\rg \mu \ll 1$. The relative difference exceeds $10\%$ when $\rg \mu=0.4$.

In the following, we refer to $N_{nlm} \equiv M_\mathrm{s}/\omega_{nlm}$ as the {\it particle number} and use it to rescale all relevant quantities, with the choice $M_\mathrm{s} = \omega_{011}$. Quantities proportional to $M_\mathrm{s}$ can then be obtained by multiplying the values obtained below with the particle number $N_{nlm}$. After the condensate reaches the superradiance threshold and before the timescale of GW emission, the condensate mass in the non-relativistic limit is approximately \cite{Brito:2017zvb,Guo:2022mpr}
\begin{align}
    M_\mathrm{s} \approx r_{g0}\mu (a_{*0}-4 r_{g0}\mu) M_0,
\end{align}
where the subscript ``0'' denotes the initial value of the corresponding quantity. Accordingly, the particle number can be estimated as
\begin{align}
    N_{011} = \frac{M_\mathrm{s}}{\omega_{011}} \approx 8.35\times10^{76} \left(\frac{M_0}{10M_\odot}\right)^2 \frac{a_{*0}-4 r_{g0}\mu}{0.1},
\end{align}
where $M_\odot$ is the solar mass.

\section{Group Velocity Distribution}
\label{sec:Rotation}

As discussed in the introduction, the group velocity distribution of a superradiant scalar field must be very different from that of the phase velocity, otherwise the motion of the condensate would exceed the speed of light far from the BH. Taking $\rg\mu=0.1$ for instance, the Bohr radius of the condensate is $r_\mathrm{b} = 100 r_g$. At this radius, the phase velocity is $v \sim \mu r_\mathrm{b} = 10 \gg 1$. 

\subsection{Energy density}
\begin{figure}
	\centering 
	\includegraphics[width=0.48\textwidth]{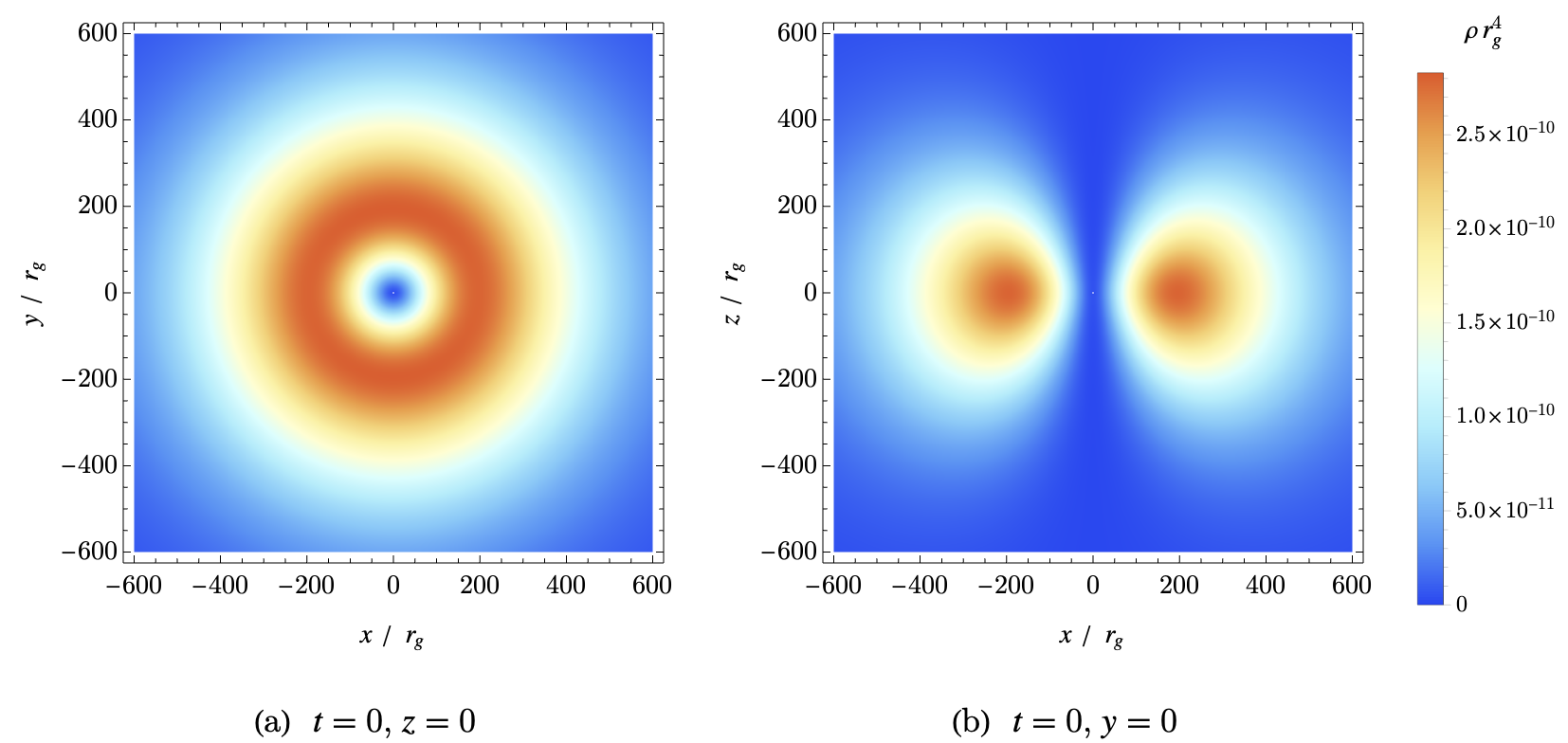}	
	\caption{The distribution of the energy density $\rho$ in the $z = 0$ (panel (a)) and $y = 0$ (panel (b)) planes. Only the $\{0,1,1\}$ mode is included in the condensate. The scalar field is normalized such that $M_\mathrm{s} = \omega_{011}$. Here, we choose $\rg\mu = 0.1$ and $a_* \approx 0.38$, to satisfy the superradiant threshold $\omega_{011} = \Omega_\mathrm{H}$.} 
	\label{fig:rho}
\end{figure}

What is physical is the energy and momentum current. Figure~\ref{fig:rho} shows the distribution of the energy density in the $z = 0$ and $y = 0$ planes for $\rg\mu=0.1$. The energy density reaches its maximum at $r_\mathrm{p} = 194.53\rg \approx 2r_\mathrm{b}$ on the equatorial plane, with a maximum value of $2.83 \times 10^{-10}\ \rg^{-4}$. These values can be converted to SI units using
\begin{align}
    r_g \approx 1.48\times10^4\ \mathrm{m} \cdot \left(\frac{M}{10M_\odot}\right),
\end{align}
and
\begin{align}
    r_g^{-4} \approx 4.15 \times10^{-39}\ \mathrm{GeV}/\mathrm{cm}^3\cdot \left(\frac{10M_\odot}{M}\right)^4.
\end{align}
Assuming a BH with mass $M=10 M_\odot$ and a condensate with the particle number $N_{011} = 10^{77}$, the maximum energy density is approximately $1.17\times10^{29}\ \mathrm{GeV}/\mathrm{cm}^3$, which is many orders of magnitude larger than the local dark matter halo density of $0.39\ \mathrm{GeV}/\mathrm{cm}^3$ in the Milky Way \cite{Catena:2009mf}.

\begin{figure*}[!ht]
	\centering 
	\includegraphics[width=0.98\textwidth]{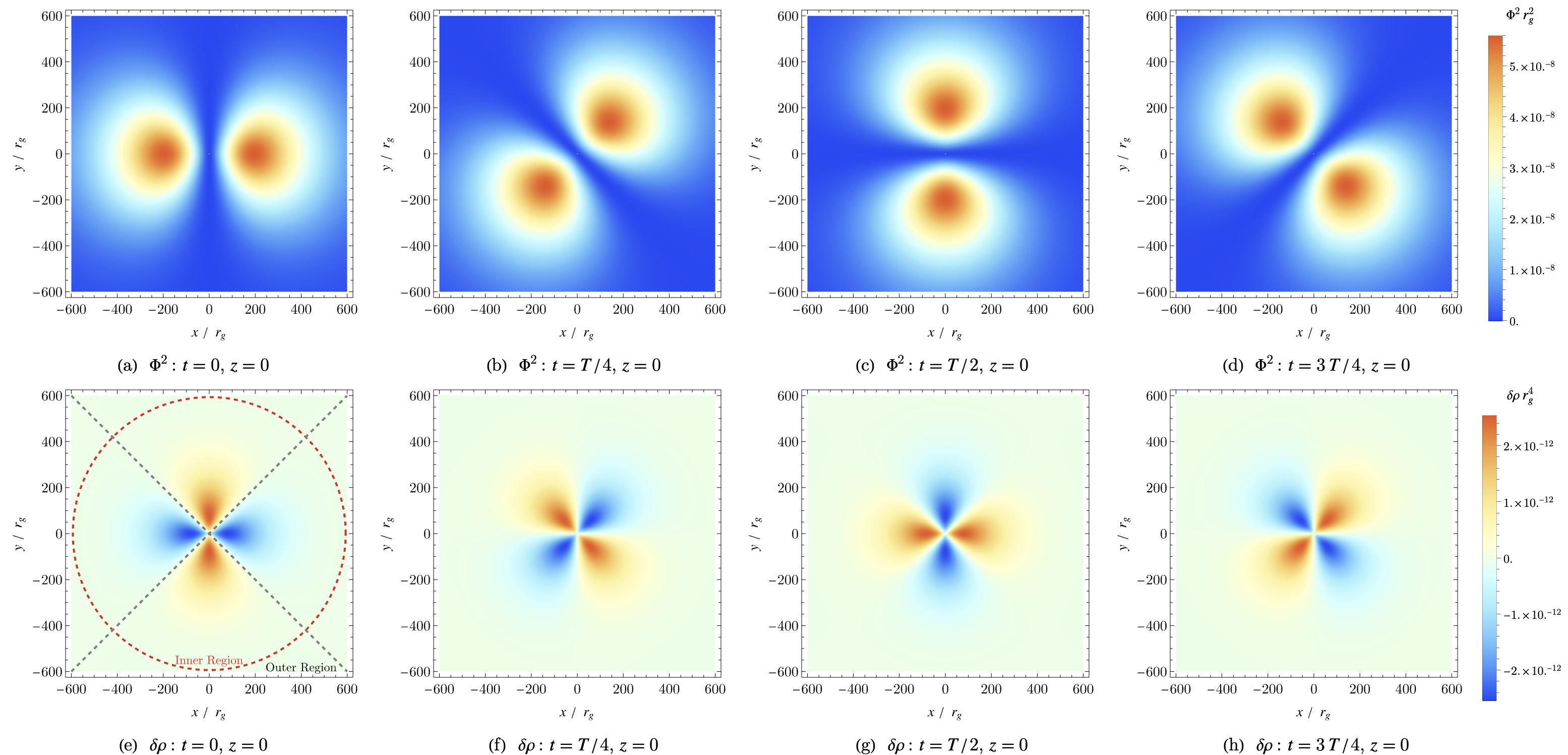}
	\caption{Distributions of the scalar field squared $\Phi^2$ (first row) and the energy density difference $\delta\rho \equiv \rho - \rho_\mathrm{ave}$ (second row) at different times in the $z = 0$ plane. The columns from left to right denote the times from $t=0$ to $3T/4$. In panel (e), surfaces with $\delta \rho = 0$ are plotted as dashed gray lines and the red circle, dividing the entire space into eight regions. Other parameters are the same as in Fig.~\ref{fig:rho}.} 
	\label{fig:delta_rho_and_Phi_square_xy}
\end{figure*}

\begin{figure*}[!ht]
	\centering 
	\includegraphics[width=0.98\textwidth]{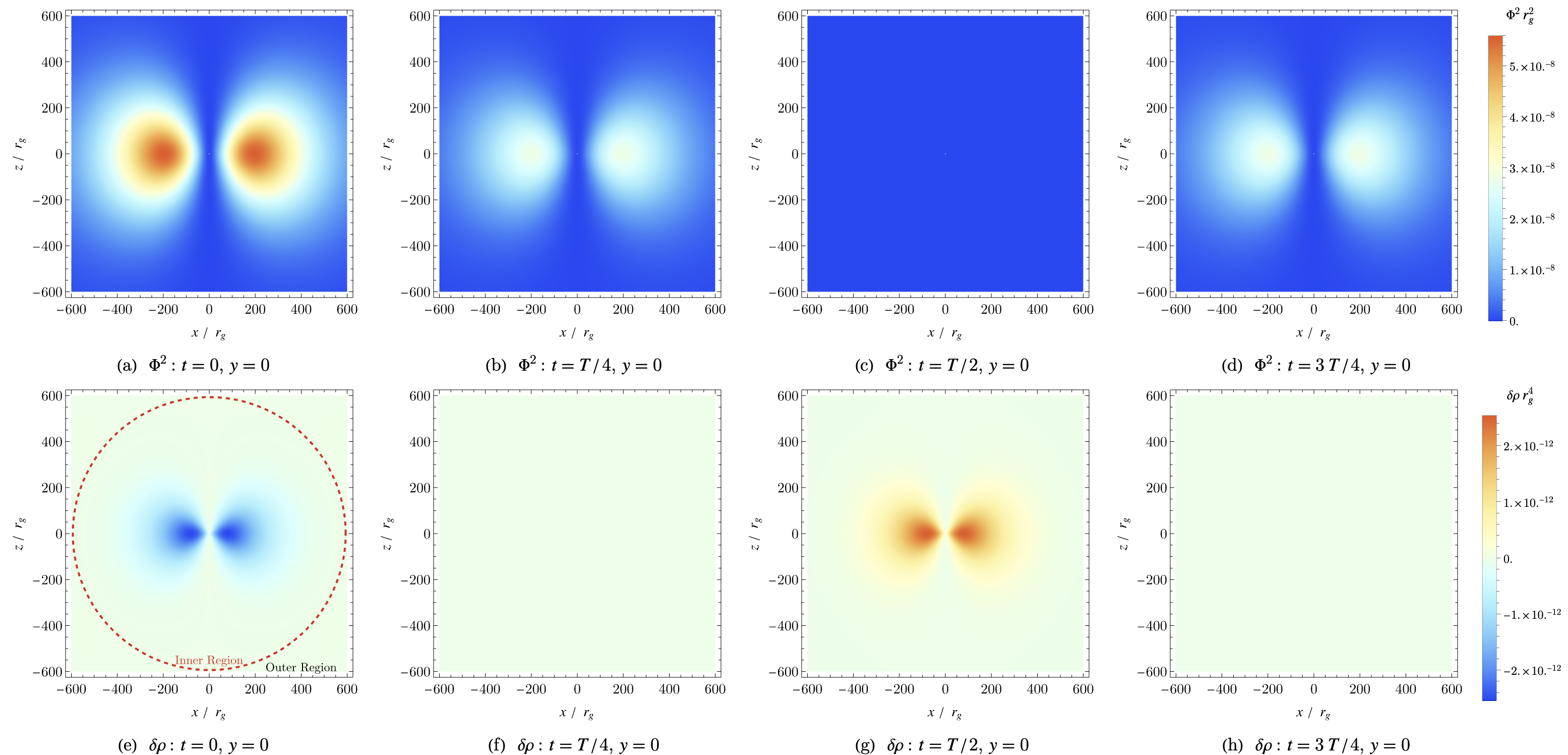}
	\caption{Same as Fig.~\ref{fig:delta_rho_and_Phi_square_xy}, but in the $y=0$ plane.} 
	\label{fig:delta_rho_and_Phi_square_xz}
\end{figure*}

Interestingly, the energy density is nearly axisymmetric. To reveal its non-axisymmetry, one first calculates the average energy density at each point in space by
\begin{align}
    \rho_\mathrm{ave}(\vec x) = \frac{1}{T} \int_{0}^{T} \rho(t,\vec x) dt,
\end{align}
where $T\equiv \pi/\omega_{011}$ denotes the variation period of the energy density. The energy density difference $\delta\rho(t,\vec x) \equiv \rho(t,\vec x) - \rho_\mathrm{ave}(\vec x)$ presents a quadrupolar structure, although the scalar field squared $\Phi^2$ exhibits a dipole. The distributions of $\Phi^2$ and $\delta\rho$ on the $z=0$ and $y=0$ planes are compared in Figs.~\ref{fig:delta_rho_and_Phi_square_xy} and \ref{fig:delta_rho_and_Phi_square_xz}, respectively. Panels in different columns indicate the distributions at different times, from $t=0$ to $3T/4$. When $t=0$, $\delta\rho$ reaches its maximum value $\delta \rho_\mathrm{max} = 2.54\times10^{-12}\,\rg^{-4}$ at $x=z=0$ and $y =\pm 79.92\,\rg$. Its minimum value is $-\delta \rho_\mathrm{max}$, located at $y=z=0$ and $x =\pm 79.92\,\rg$ when $t=0$. The relative energy density ratio $|\delta \rho/\rho_\mathrm{ave}|$ decreases monotonically as the radial distance increases. For example, along the $x$-axis in the panel (e), the ratio decreases from $0.48$ near the horizon $r=1.92\,\rg$ to $0.01$ at $r=119.40\,\rg$.

\begin{figure}[t]
	\centering 
	\includegraphics[width=0.4\textwidth]{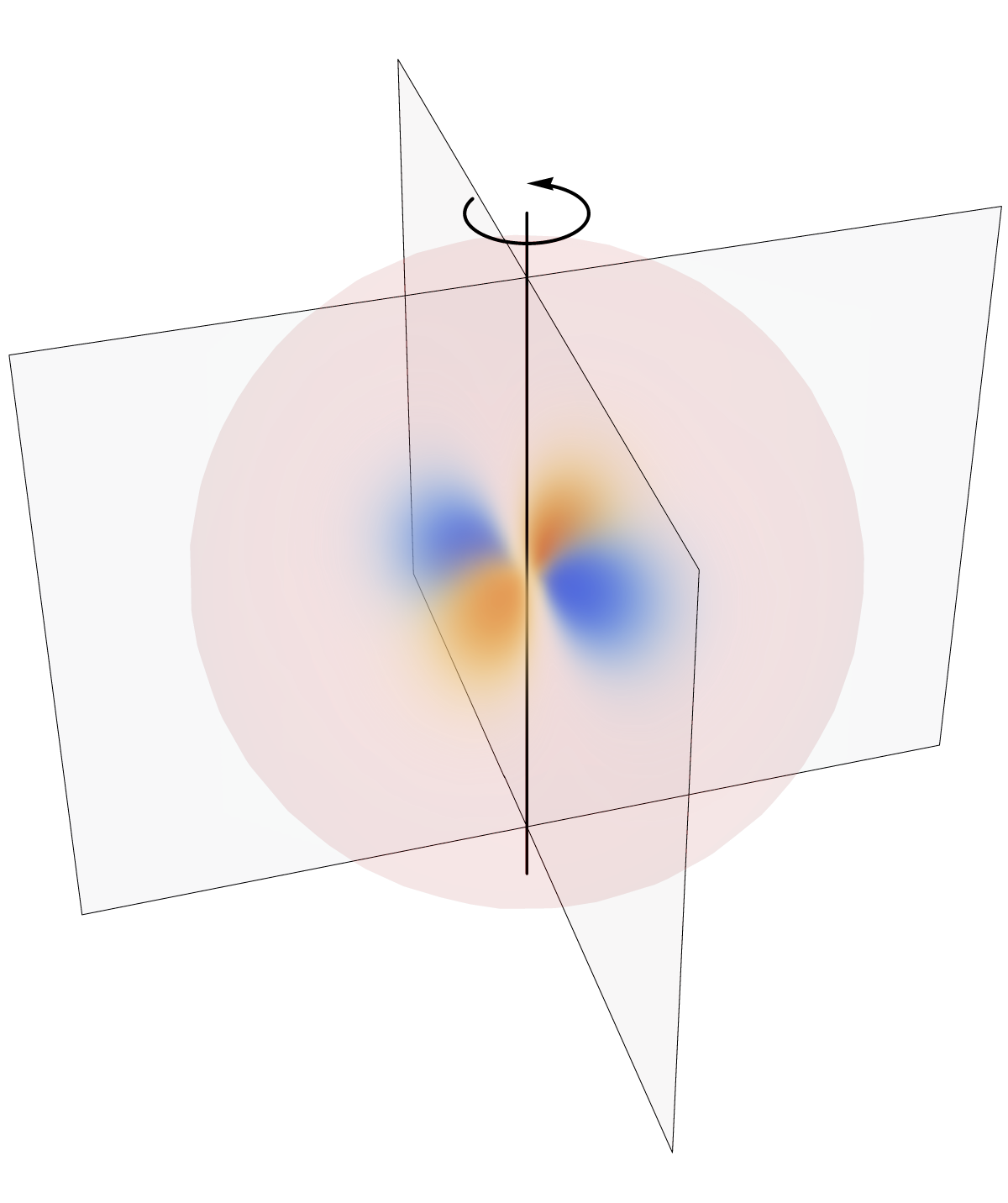}
	\caption{Schematic diagram of the surfaces where the energy density difference satisfies $\delta \rho = 0$. These surfaces are shown by two gray planes and a red sphere. The distribution of $\delta \rho$, the rotation axis, and the direction of rotation are also depicted.} 
	\label{fig:3D}
\end{figure}

Several notable features can be observed in Figs.~\ref{fig:delta_rho_and_Phi_square_xy} and \ref{fig:delta_rho_and_Phi_square_xz}. Regions where $\Phi^2$ concentrates have negative values of $\delta\rho$. Regions with positive and negative $\delta\rho$ alternate along the $\varphi$ direction, suggesting the existence of two surfaces where $\delta \rho = 0$ between these regions, shown by gray dashed lines in Fig.~\ref{fig:delta_rho_and_Phi_square_xy}(e) and gary planes in Fig.~\ref{fig:3D}. An additional $\delta \rho = 0$ surface is shown by the red dashed circle in Figs.~\ref{fig:delta_rho_and_Phi_square_xy}(e) and \ref{fig:delta_rho_and_Phi_square_xz}(e), and the red sphere in Fig.~\ref{fig:3D}. These surfaces divide the entire space into eight regions. For convenience,  we refer to the regions separated by the red sphere as the {\it inner} and {\it outer regions}. The regions with positive (negative) $\delta \rho$ are hereafter referred to as the positive (negative) regions. The red sphere indicates the presence of a negative (positive) outer region outside a positive (negative) inner region, which is too weak to be clearly visible in Figs.~\ref{fig:delta_rho_and_Phi_square_xy} - \ref{fig:3D}.

The BH, the quadrupolar, and the scalar field squared all rotate in the same direction, as indicated by the arrow in Fig.~\ref{fig:3D}. 
In Fig.~\ref{fig:delta_rho_and_Phi_square_xy}, the rotation is counter-clockwise. In Fig.~\ref{fig:delta_rho_and_Phi_square_xz}, the left half panel moves outwards while the right half moves inwards. Their configurations undergo a $180^\circ$ rotation from $t = 0$ to $T$ in the azimuthal direction, resembling the motion of a rigid body.

\subsection{Velocity}

\begin{figure}[t]
	\centering 
	\includegraphics[width=0.48\textwidth]{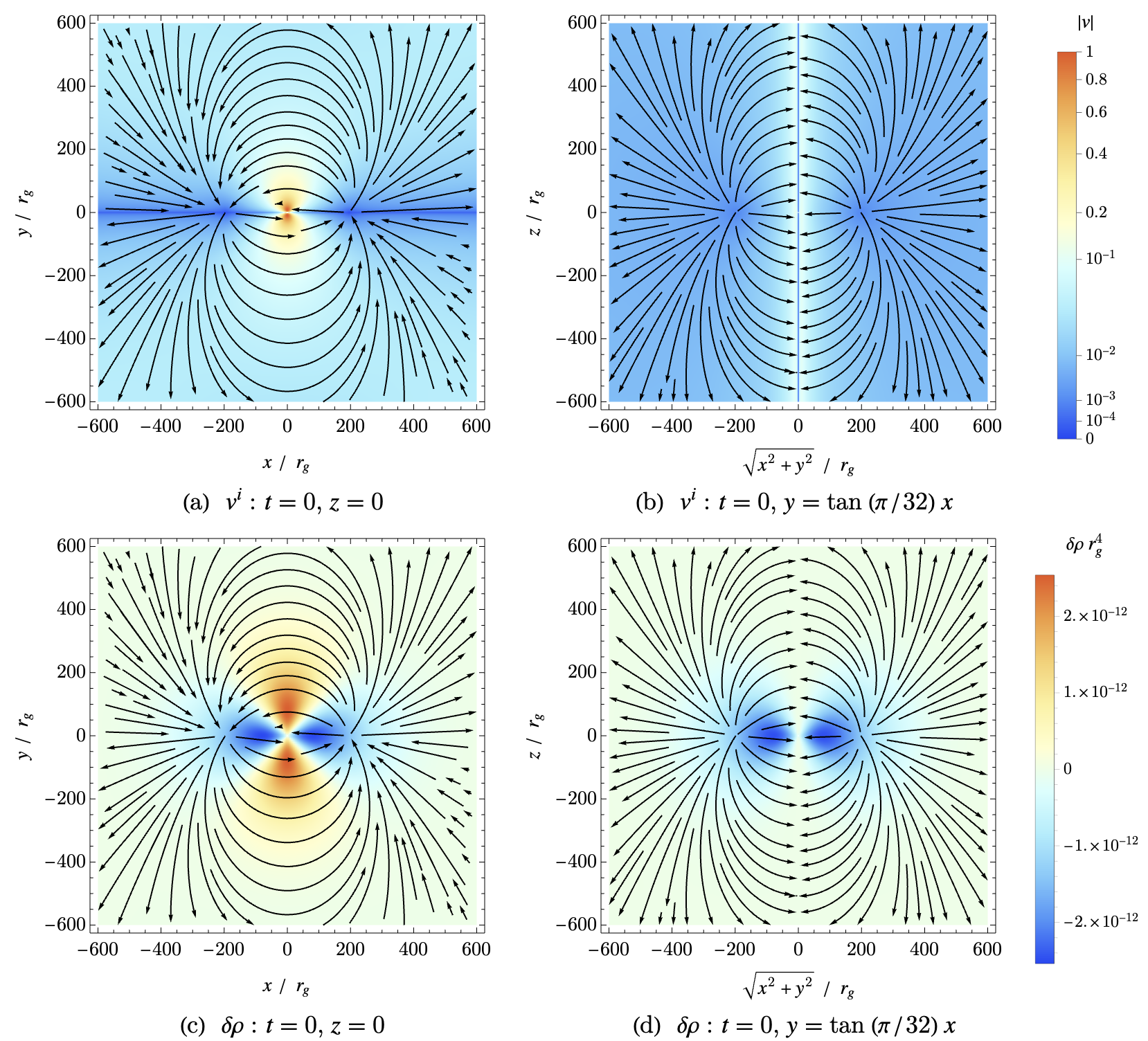}	
	\caption{Distributions of the velocity $v^{i} = P^{i}/\rho$ (first row) and energy density difference $\delta\rho \equiv \rho - \rho_\mathrm{ave}$ (second row) at $t=0$ in the $z = 0$ (first column) and $y = \tan(\pi/32)x$ (second column) planes. The arrows in all planes denote the velocity projections onto the corresponding planes. Other parameters are the same as in Fig.~\ref{fig:rho}.
    }
	\label{fig:rotation}
\end{figure}

In fact, the apparent rotation of the condensate is a collective effect resulting from local motions with radial, polar, and azimuthal components. The local group velocity is defined as
\begin{align}\label{eq:velocity}
    v^{i} = P^{i} / \rho,
\end{align}
where the momentum density $P^{i}$ is given by
\begin{align}\label{eq:momentum_density}
    P^{i} \equiv T^{\nu}_{\ \sigma} \left(e_{t}\right)^{\sigma} \left(e^{i}\right)_{\nu}.
\end{align}
Figure~\ref{fig:rotation}(a) presents the group velocity distributions on the $z = 0$ plane at $t = 0$. The color indicates the magnitude of the velocity, while arrows denote the velocity projections onto the $z=0$ plane. For convenience, Fig.~\ref{fig:rotation}(c) shows the distribution of $\delta\rho$, overlapped with the projected velocity direction. At $t=0$, the energy flux vanishes on the $y=0$ plane, which separates the space into two regions. In the first (third) quadrant, some streamlines from the inner region extend to the outer region, while others flow into the second (fourth) quadrant. As a result, energy accumulates in the second and fourth quadrants, while dissipates from the first and third quadrants. Altogether, it produces the counter-clockwise apparent rotation of $\delta\rho$ in the $x-y$ plane. Accordingly, the zero-flux plane, which is $y=0$ at $t=0$, rotates in the same pace as $\delta\rho$.

Besides the motion in the transverse directions, the energy flow also has non-zero $z$ component. Figure \ref{fig:rotation}(b) shows the magnitude of the velocity on the $y = \tan(\pi/32)x$ plane as well as its direction projection on this plane. Combining Fig.~\ref{fig:rotation}(a,b), the three-dimensional distribution of streamlines with $y>0$ are similar to the electric field lines of an electric dipole with positive and negative charges located at $(r,\theta,\varphi) \approx (200\,\rg,\pi/2,0)$ and $(200\,\rg,\pi/2,\pi)$, respectively. On the other hand, the velocity steamlines in the $y<0$ region are similar to the electric field lines of a reversed dipole. Then these two regions are glued at $y=0$ plane, on which the energy flux vanishes. This picture is qualitatively corerect except in the region $r \gg r_\mathrm{p}$, which will be explained later.

\begin{figure}[ht]
	\centering 
	\includegraphics[width=0.45\textwidth]{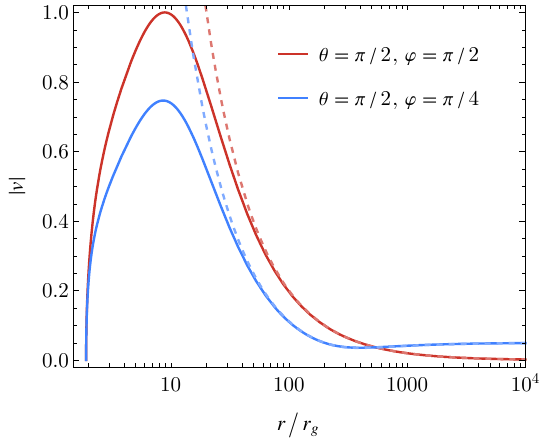}
    \caption{Velocity as a function of radial distance for two representative cases at $t = 0$, along with their corresponding asymptotic behaviors. The solid red curve represents the velocity along $y$-axis in Fig.~\ref{fig:rotation} (i.e., the angular direction defined by $\theta = \pi/2$ and $\varphi = \pi/2$), while the solid blue curve corresponds to the angular direction defined by $\theta = \pi/2$ and $\varphi = \pi/4$. The light red and light blue dashed curves indicate the asymptotic behaviors are given by Eqs.~\eqref{eq:v_asym}.
    }
	\label{fig:Velocity}
\end{figure}

From the symmetry argument, the energy flux is purely azimuthal on the $x=0$ plane at $t=0$. To show there is no superluminal motion, the velocity magnitude $|v| \equiv \sqrt{v^i v_i}$ along the positive $y$ axis as a function of radial distance $r$ is plotted with the solid red curve in Fig.~\ref{fig:Velocity}. As $r$ increases, $|v|$ first increases, then decreases to zero at infinity, peaking at $r = 8.90\,\rg$ where $|v| \sim 1$. The cruves along other directions exhibit a little more complex behavior. In Fig.~\ref{fig:Velocity}, the $|v|$ in $\theta=\pi/2$ and $\varphi=\pi/4$ direction is plotted with the solid blue curve. The velocity magnitude initially exhibits a similar trend with a smaller peak of $0.747$ at $r = 8.63\,\rg$. It then reaches its minimum value of 0.0358 at $r = 398.05\,\rg$, and subsequently increases, approaching a constant value of 0.0502 at infinity. Therefore, the condensate moves relativistically only in the region $r\lesssim 50 r_g$. At the Bohr radius $r_\text{b}$, the motion can be safely taken as non-relativistic, which qualifies the perturbative treatment in Ref.~\cite{Baumann:2018vus}.

For deeper understanding of the group velocity distribution, we analyze the system in the non-relativistic limit, $\rg \mu \ll 1$. Focusing on the region where $r \gg \rg$, the radial equation \eqref{eq:radial_equation} has an asymptotic solution~\cite{Detweiler:1980uk,Bao:2022hew}:
\begin{align}
    R_{nl}(r) \sim (-2qr)^{l} e^{qr} U(-n,2l+2,-2qr),
\end{align}
where $U$ is the confluent hypergeometric function of the second kind. The spheroidal harmonic $S_{nlm}(\theta)$ reduces to the associated Legendre function $\mathcal{P}_{lm}(\theta)$. In flat spacetime, the energy and momentum densities of the $\{0,1,1\}$ mode are given by
\begin{subequations}
    \begin{align}
        \rho &\approx \omega_{011} R^2_{01}(r) \mathcal{P}^2_{11}(\theta ),
        \\
        P^{r} &=  R_{01}(r) R_{01}'(r) \mathcal{P}^2_{11}(\theta ) \sin \left(2 \omega_{011} t - 2 \varphi\right),
        \\
        P^{\theta} &=  \frac{R^2_{01}(r) \mathcal{P}_{11}(\theta) \mathcal{P}_{11}'(\theta) \sin \left(2 \omega_{011} t - 2 \varphi\right)}{r},
        \\
        P^{\varphi} &=  \frac{2 R^2_{01}(r) \mathcal{P}^2_{11}(\theta ) \sin^2\left(\omega_{011} t - \varphi \right)}{r \sin\theta}.
    \end{align}
\end{subequations}
Substituting the asymptotic expressions of $R_{01}(r)\mathcal{P}_{11}(\theta)$, the velocity at large distances can be approximated as
\begin{subequations}
    \label{eq:v_asym}
    \begin{align}
        v^{r} &\approx  \frac{1 - r/r_\mathrm{p}}{\mu r} \sin \left(2 \omega_{011} t - 2 \varphi\right),
        \\
        v^{\theta} &\approx  \frac{\cot \theta }{\mu r} \sin \left(2 \omega_{011} t - 2 \varphi\right),
        \\
        v^{\varphi} &\approx  \frac{2}{\mu r \sin\theta} \sin^2 \left(\omega_{011} t - \varphi\right),
    \end{align}
\end{subequations}
where the approximations $\omega_{011}\approx\mu$ and $r_\mathrm{p}\approx 2 r_\mathrm{b}$ are used. The asymptotic expression for $v^{r}$ ($v^{\theta}$ and $v^{\varphi}$) is valid when $\mu r\gg 1$ ($\mu r\sin\theta \gg 1$). They are plotted as the dashed curves and compared with the full result at $t=0$ in Fig.~\ref{fig:Velocity}. The agreement is excellent for $r\gtrsim r_\text{b}=100\,\rg$. Specifically, the asymptotic formulae are accurate to within $10\%$ for $r \gtrsim 30\,r_g$.

The asymptotic velocity components in Eqs.~\eqref{eq:v_asym} share the same period as $\delta\rho$. This demonstrates that the condensate's apparent rotation arises from a collective effect of local motions. The polar and azimuthal velocities decrease proportionally to $1/r$ with increasing radial distance. The direction of $v^{\theta}$ reverses across the equatorial plane, while the sign of $v^{\varphi}$ is the same in the entire space, aligned with the BH spin. The radial velocity component $v^{r}$ changes sign at $r = r_\mathrm{p}$. These observations are consistent with Fig.~\ref{fig:rotation}. 


We further focus on the region where $r \gg r_\mathrm{p}$. With $\omega_{011}t -\varphi = k\pi$ and $k\in \mathbb{Z}$, all velocity components vanish, defining the zero-flux plane. With $\omega_{011}t -\varphi = (k+1/2)\pi$ and $k\in \mathbb{Z}$, the flux is purely in the azimuthal direction. Otherwise, the flux is almost purely radial at $r\to \infty$, with a non-zero velocity approaching $(1 / r_\mathrm{p} \mu)\sin(2\omega_{011}t-2\varphi) \approx (\mu \rg / 2) \sin(2\omega_{011}t-2\varphi)\ll 1$. At $t=0$, using the same parameters of the blue curve in Fig.~\ref{fig:Velocity}, it gives $0.05$, only slightly smaller than the numerical result $0.0502$. 


With $r \gg r_\mathrm{p}$, the radial component of the velocity dominantes except close to the planes defined by $\omega_{011}t-\varphi = k\pi$ or $(k+1/2)\pi$, where $k\in\mathbb{Z}$. This behavior is markedly different from the dipole electric field lines, which always wind towards the negative charge no matter how far they are from the dipole. As the time $t$ increases, these two planes rotate together with the field $\Phi$. The energy at $r\gg r_\mathrm{p}$ flows inwards and outwards alternately along the radial direction with period $\pi/\omega_{011}$.

Finally, we discuss the velocity behavior as $z$ varies. When $\mu z\gg 1$ and $\mu \sqrt{x^2+y^2} \gg 1$, the velocity becomes
\begin{subequations}
    \begin{align}
        v^{x} &\approx  \frac{\sin (2 \omega_{011} t - \varphi)-\sin (\varphi )}{\mu\sqrt{x^2+y^2}},
        \\
        v^{y} &\approx -\frac{\cos (2 \omega_{011} t - \varphi)-\cos (\varphi )}{\mu\sqrt{x^2+y^2}},
        \\
        v^{z} &\approx  -\frac{\sin (2 \omega_{011} t - 2\varphi)}{\mu r_\mathrm{p}}.
    \end{align}
\end{subequations}
These asymptotic expressions indicate that the velocity is approximately a constant as $z$ varies when $t$, $x$ and $y$ are fixed, which is consistent with Fig.~\ref{fig:rotation}(b). These asymptotic expressions agree with numerical result as well. For example, at $t=0$, $\sqrt{x^2+y^2}=200\,\rg$ and $\varphi = \pi/32$ in Fig.~\ref{fig:rotation}(b), the numerical result of the constant yields $|v| \approx 1.378\times10^{-2}$ while the asymptotic approximation gives $|v| \approx 1.383\times10^{-2}$.

\section{Summary and discussion}
\label{sec:Summary}
In this work, we performed a detailed investigation of the local group velocity distribution of superradiant condensates around rotating BHs, focusing on real scalar fields without self-interactions or couplings to other fields. Although the phase propagation occurs only in the azimuthal direction, we demonstrated that the apparent rotation of the condensate is a collective effect arising from local motions in all three directions, by analyzing the spatial distribution and temporal evolution of the energy density and velocity.


We first briefly reviewed scalar superradiance and the BH-condensate system in Sec.~\ref{sec:BHCondensate}. Then, we discussed the normalization of the scalar field in Sec.~\ref{sec:normalization}. We explicitly showed the squared distribution function $f_{lm}^2$ in Eq.~\eqref{eq:real_scalar} is not equal to $M_\mathrm{s}/\omega_{nlm}$ in curved spacetime when the scalar field satisfies the condition \eqref{eq:normalization}.

In Sec.~\ref{sec:Rotation}, we demonstrated that scalar field squared $\Phi^2$ exhibits a dipole, while the energy density $\rho$ of the condensate is approximately axisymmetric and peaks at $r_\mathrm{p} \approx 2r_\mathrm{b}$. To reveal the quadrupole structure of $\rho$, we defined the energy density difference $\delta \rho (t,\vec x) \equiv \rho (t,\vec x) - \rho_\mathrm{ave}(\vec x)$, where $\rho_\mathrm{ave}(\vec x)$ denotes the time average of $\rho$ over one period at location $\vec x$. The surfaces where $\delta \rho = 0$ divide the entire space into eight distinct inner and outer regions, as illustrated in Fig.~\ref{fig:3D}. Interestingly, the region where $\Phi^2$ concentrates has negative values of $\delta\rho$.

Through detailed numerical and analytical studies, we found that the apparent rotation of the condensate is actually a collective effect. In the region $r \lesssim r_\mathrm{p}$ at $t=0$, the group velocity streamline distribution in the $y>0$ region resembles the electric field lines of an electric dipole, with positive and negative charges located at $(r,\theta,\varphi) \approx (200\,\rg,\pi/2,0)$ and $(200\,\rg,\pi/2,\pi)$, respectively. The group velocity steamlines in the $y<0$ region resemble the electric field lines of a reversed dipole. These two regions are glued at $y=0$ plane, on which the energy flux vanishes. As the time $t$ increases, the velocity streamlines rotate with period $\pi/\omega_{011}$.

In the region $r \gg r_\mathrm{p}$, this analogy no longer holds. The radial group velocity dominates in this region, except close to the planes defined by $\omega_{011}t-\varphi = k\pi$ or $(k+1/2)\pi$, with $k\in\mathbb{Z}$. This behavior is markedly different from the dipole electric field lines, which always wind towards the negative charge even at large distances. As the time $t$ increases, these two planes rotate together with the field $\Phi$. The energy at $r\gg r_\mathrm{p}$ flows inwards and outwards alternately along the radial direction with period $\pi/\omega_{011}$.


These conclusions follow from the asymptotic forms of the velocity within the non-relativistic approximation. We showed that the magnitudes of the polar and azimuthal velocities decrease as $1/r$ when $\mu r \sin\theta \gg 1$. The radial velocity exhibits a similar behavior when $1/\mu \ll r \ll r_\mathrm{p}$ and approaches $ (\rg\mu/2) \sin (2\omega t-2 \varphi)$ when $r \gg r_\mathrm{p}$. These asymptotic behaviors confirm that the motion remains non-relativistic at large distances and there is no superluminal motion. Additionally, the velocity as a function of $z$ approaches a constant at large $z$ with $t$, $x$, and $y$ fixed.

Our findings offer a refined understanding of the apparent rotation of superradiant condensates. This work lays a foundation for future phenomenological studies, such as photon birefringence and dynamical friction of objects moving inside a superradiant scalar field.

\section*{Acknowledgements}
We thank the referees for the questions about possible observables of the group velocity distribution.
This work is supported by the National Natural Science Foundation of China (Grants Nos. 12447105, 12075136 and 124B2098) and the Natural Science Foundation of Shandong Province (Grant No. ZR2020MA094)




\end{document}